\documentstyle[11pt,epsf]{article}
\def\andvol#1{{\bf \@JLone<#1>} (\@JLtwo<#1>), \@Jpage<#1>}

\title{
Thermal Relaxation in One-Dimensional Self-Gravitating Systems
}

\author{
Junichiro  {\sc Makino}\footnote{E-mail address: makino@grape.c.u-tokyo.ac.jp} 
\\
Department of Systems Science,
College of Arts and Sciences,\\
University of Tokyo, Tokyo 153-8902}

\begin{document}

\maketitle

\begin{abstract}
In this paper, we study the thermal relaxation in the
one-dimensional self-gravitating system, or the so-called sheet model. 
According to the standard argument, the thermal relaxation time of the
system is around $Nt_c$, where $N$ is the number of sheets and $t_c$
is the crossing time. It has been claimed that the system does not
reach the thermal equilibrium in this thermal relaxation timescale,
and that it takes much longer time for the system to reach true
thermal equilibrium.

We demonstrate that this behavior is explained simply by the
fact that  the relaxation time is long. The relaxation time of sheets
with average binding energy is $\sim 20 Nt_c$, and that of sheets with 
high energy can exceed $1000 Nt_c$.  Thus,
one needs to take the average over the relaxation timescale of high-energy
sheets, if one wants to look at the thermal characteristic of these
high energy sheets. 
\end{abstract}

\section{Introduction}

The one-dimensional self-gravitating many-body system was originally
discussed mainly as a simple toy model to understand the violent
relaxation, \cite{Lynden-Bell1967}\ because the thermal relaxation
timescale of its discrete realization, the sheet model, was believed
to be long. Until 1980s, it had been generally accepted that the
thermal relaxation time of the system of $N$ equal-mass sheets is of
the order of $N^2t_c$, where $t_c$ is the crossing time of the
system.

However, by means of numerical simulation Luwel {\it et
al.}\cite{Luwel1984} have demonstrated that the relaxation time is of
the order of $Nt_c$. Reidel and Miller \cite{Reidl1987,Reidl1988}
reached a similar conclusion, though they reported the presence of
systems which apparently did not relax for much longer timescale. 

In a series of papers, Tsuchiya {\it et al.}
\cite{Tsuchiya1994,Tsuchiya1996,Tsuchiya1997} have studied the thermal 
relaxation process of one-dimensional self-gravitating systems in
detail, by means of the numerical integration over very long timescale
(some of their experiments covered $5\times 10^8 t_c$). They claimed
that  the thermal relaxation of the sheet model proceeds in a
highly complex manner. In the ``microscopic relaxation timescale'' of
$Nt_c$, each sheet forgets its initial condition, and the system is
well mixed. However, according to them, the system does not really
reach the thermal equilibrium in this timescale, and the distribution
function remains different from that of the isothermal state. They
called this state a quasiequilibrium

By pursuing the time integration for much longer timescale, Tsuchiya
{\it et al.} \cite{Tsuchiya1997} found that the system exhibits the
transition from one quasiequilibrium to another, and they claimed that 
the thermal equilibrium is only realized by averaging over the
timescale longer than the timescale of these transitions. Thus, they
argued that there exists the timescale for ``macroscopic'' relaxation, 
which is much longer than the usual thermal relaxation (what they
called ``microscopic relaxation''). 

In this paper, we try to examine the nature of this ``macroscopic''
relaxation of the one-dimensional sheet model. In section 2,
we describe the numerical model. In section 3, 
we present the result of the measurement of the relaxation time. It is 
shown that the relaxation time, defined as the timescale in which 
individual sheets change their energies, depends very strongly on the 
energy itself, and is very long for high energy sheets.
This strong dependence of the relaxation timescale on the energy
naturally explains the apparent ``transient'' phenomena observed by
Tsuchiya {\it et al.}\cite{Tsuchiya1997} \  Section 4 discusses the
implication and relevance of our results.

\section{The Model}

\subsection{Sheet model}

The Hamiltonian of the sheet model is given by
\begin{equation}
H = {m \over 2}\sum_{i=1}^N v_i^2 +2\pi Gm^2\sum_{i<j}|x_i-x_j|,
\end{equation}
where $x_i$ and $v_i$ are the position and velocity of sheet $i$, $m$
is the mass of the sheets, $N$ is the number of the sheets and $G$ is
the gravitational constant. The crossing time is defined as
\begin{equation}
t_c = {1 \over 4\pi GM}\sqrt{{4E \over M}},
\end{equation}
where $M=mN$ is the total mass of the system.  Following Tsuchiya {\it 
et al.}\cite{Tsuchiya1997} and others, we use
the system of units in which $M=4E=4\pi G = 1$. In this system, $t_c =
1$.

A unique nature of the one-dimensional gravitational system is that
there exists the thermal equilibrium, unlike its counterpart in three
dimensions. 
Rybicki \cite{Rybicki1971} obtained the distribution function
\begin{equation}
f(\varepsilon) = {1 \over 8} \left({1 \over 2\pi}\right)^{1/2}
\left({3M \over 2E}\right)^{3/2}\exp\left(-{3M \over
2E}\varepsilon\right),
\label{eq:isotherm}
\end{equation}
where $\varepsilon$ is the specific binding energy defined as
\begin{equation}
\varepsilon = {v^2 \over 2} +\Psi(x) - \Psi(0).
\end{equation}
Here, $\Psi(x)$ is the specific  potential energy. This distribution
function satisfies the relation
\begin{equation}
\exp\left(-{3M \over 2E}\varepsilon\right)={\rm sech}^2
\left({3x \over 8E}\right).
\end{equation}

We performed the time integration of the system with $N=16$, 32, 64,
128 and 256. For all systems, the initial condition is a water-bag
with the aspect ratio $x_{max}/v_{max} = 2.5$.

\subsection{Numerical method}

The important character of the sheet model is that one can calculate
the exact orbit of each sheet until two sheets cross each other. Thus,
we can integrate the evolution of the system precisely (except for the 
round-off error).  This may
sound like a great advantage, compared to the systems in higher
dimensions whose orbits can be calculated only numerically.  Instead
of numerically integrating the orbit of each sheet, we can calculate
the exact orbit for any sheet, until it collides with the neighboring
sheet. Thus, by arranging the pairs using heap, we can handle each
collision in $\log N$ calculation cost.

Note, however, that typically each sheet collides with all other
sheets in one crossing time. Thus, the calculation cost is $O(N^2\log
N)$ per crossing time.  Our simulation with $N=64$ for $2\times 10^7
t_c$ took 8 hours on a VT-Alpha workstation with DEC Alpha 21164A CPU
running at 533 MHz. For this run, the total energy of the system was
conserved  better than $3\times 10^{-12}$.

\section{Results}

\subsection{Approach to the thermal equilibrium}

\begin{figure}
\begin{center}
\leavevmode
\epsfxsize 7cm
\epsffile{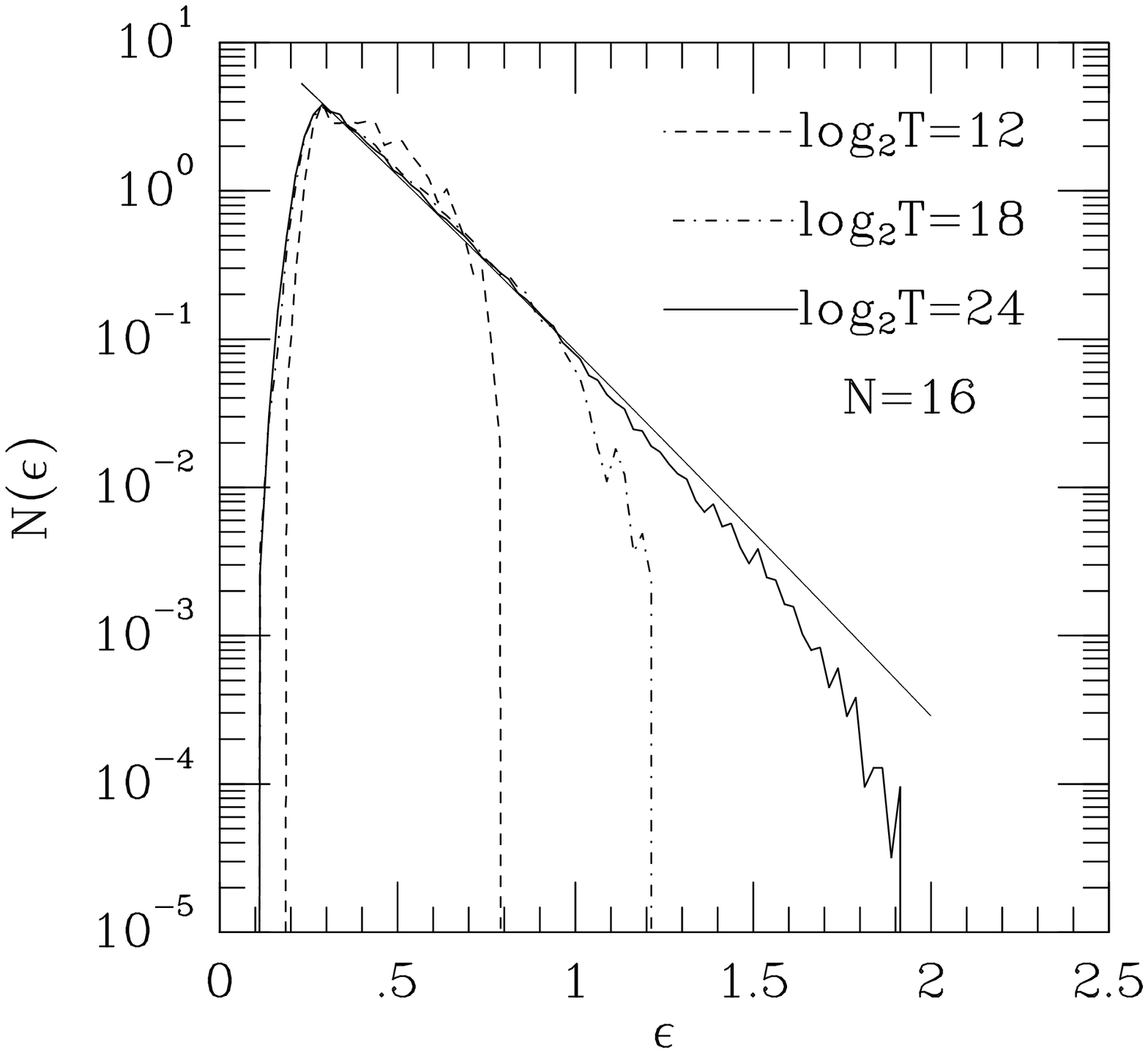} \ \ 
\epsfxsize 7cm
\epsffile{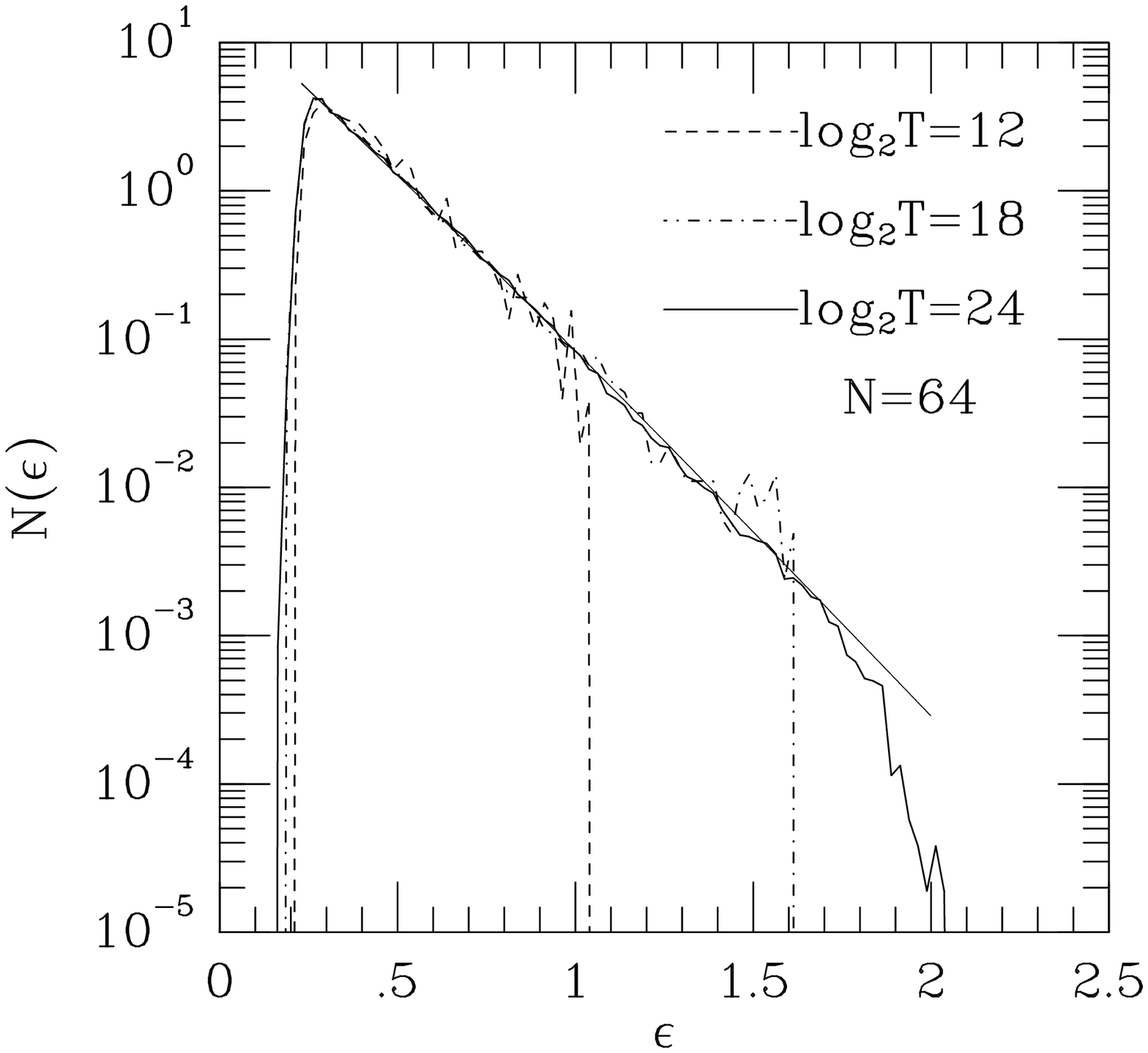}\\
\leavevmode
\epsfxsize 7cm
\epsffile{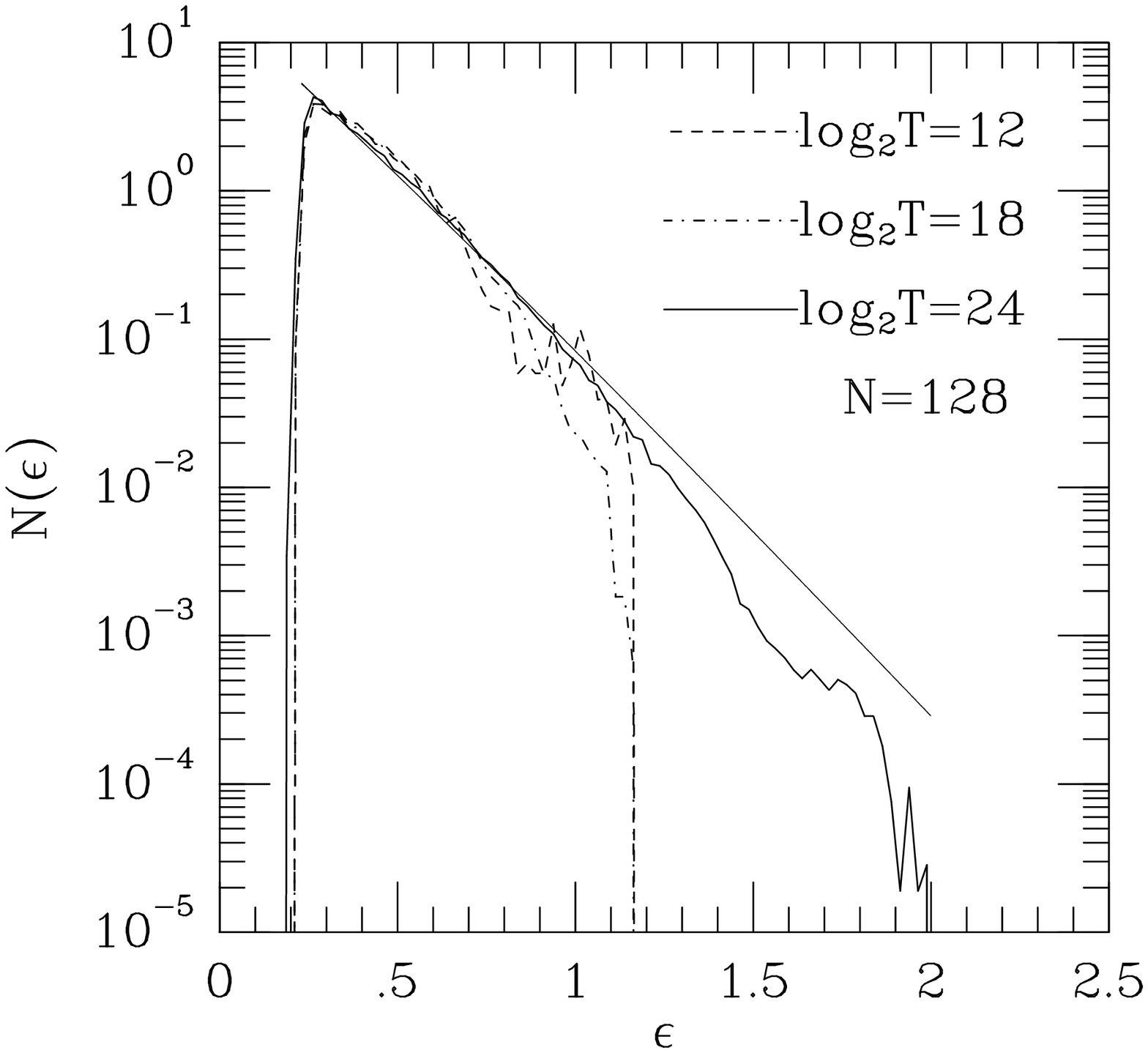}
\end{center}
\caption{The time-averaged distribution function in the energy space
$N(E)$; (a) $N=16$, (b) $N=64$, (c) $N=128$.}
\label{fig:NE}
\end{figure}

Figure \ref{fig:NE} shows the time-averaged energy distribution
function $N(\varepsilon)$, for different time periods and number of sheets. In
all figures, the thin solid curve is the energy distribution of the
isothermal distribution function of equation (\ref{eq:isotherm}). What
we see is quite clear. As we make the time interval longer, the
time-averaged distribution function approaches to the isothermal
distribution. Thus, the numerical result suggests the system is
ergodic. However, it also shows that the time needed to populate the
high-energy region is very long. The sampling time interval is 128
time units for $N=16$, and 512 time units for $N=64$ and 128. Thus, in
the case of $T=2^{18}$ and $N=16$ (dash-dotted curve in figure 1a), total
number of sample points is $2^{15}=32768$.

\begin{figure}
\begin{center}
\leavevmode
\epsfxsize 7cm
\epsffile{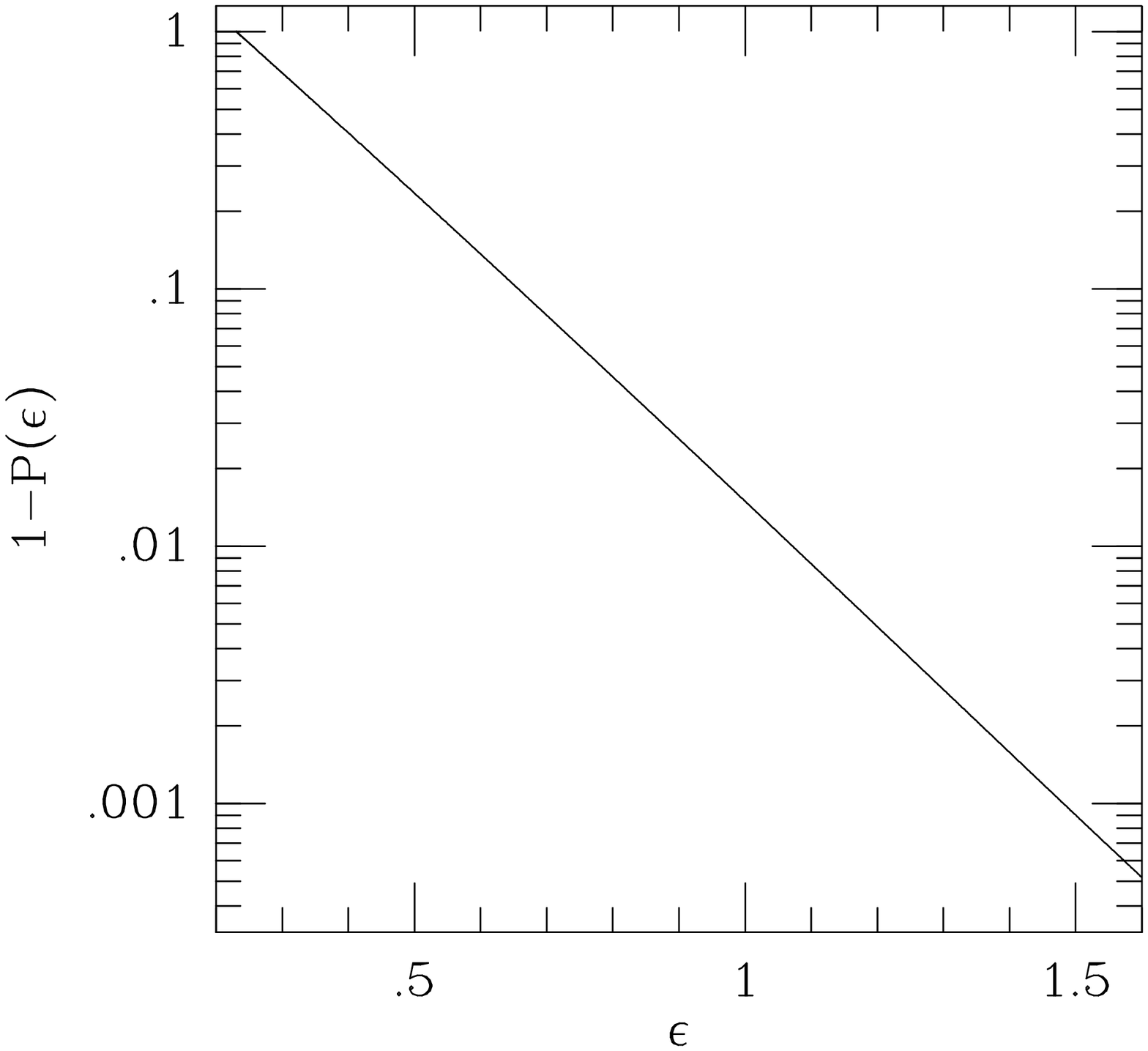}
\end{center}
\caption{The compliment of the cumulative distribution function
$1-P(e)$ for the thermal equilibrium.}
\label{fig:theory}
\end{figure}

If we can assume that the sample points are uncorrelated, the
possibility that no sample exceeds energy level $\varepsilon_0$ is
given simply by
\begin{equation}
P(\varepsilon_0,N) = [1-P(\varepsilon<\varepsilon_0)]^n,
\end{equation}
where
\begin{equation}
P(\varepsilon<\varepsilon_0) = \int_0^{\varepsilon_0} N(\varepsilon)d\varepsilon,
\end{equation}
and $n$ is the number of sample points. Figure \ref{fig:theory} shows
$1-P(\varepsilon)$ as a function of $\varepsilon$. For
$\varepsilon=1.25$, $P(\varepsilon)=0.996$, and therefore the
probability that none of 32768 samples does not exceed $\varepsilon =
1.25$ is practically zero ($<e^{-100}$). In other words, the numerical
result seems to suggest that the system is not in the thermally
relaxed state even after $2\times 10^5$ crossing times.

Of course, this result is not surprising if the relaxation time is
long.  Samples taken with the time interval shorter than the
relaxation time have a strong correlation, and therefore the effective
number of freedom can be smaller than $n$. Roughly speaking, if the
relaxation time is longer than $10^4$, our numerical result is
consistent with the assumption that the system is in the thermal
equilibrium. In the next subsection, we investigate the relaxation
time itself.

\subsection{Relaxation timescale}

We measured the  following quantities:
\begin{eqnarray}
D_1 &=& {<\varepsilon_i(t_0)-\varepsilon_i(t_0+\Delta t)> \over \Delta t},\\
D_2 &=& {<[\varepsilon_i(t_0)-\varepsilon_i(t_0+\Delta t)]^2> \over \Delta t}.
\end{eqnarray}
These quantities correspond to the coefficients of the first and
second-order terms in the Fokker-Planck equation for the distribution
function, and have been used as the measure of the relaxation in many
studies (see, {\it e.g.}, 
Hernquist and Barnes,\cite{Hernquist1990}\ Hernquist {\it et al.}
\cite{Hernquist1993}), for three-dimensional systems. However, to our
knowledge this measure has not been used for the study of the sheet
model.

\begin{figure}
\begin{center}
\leavevmode
\epsfxsize 7cm
\epsffile{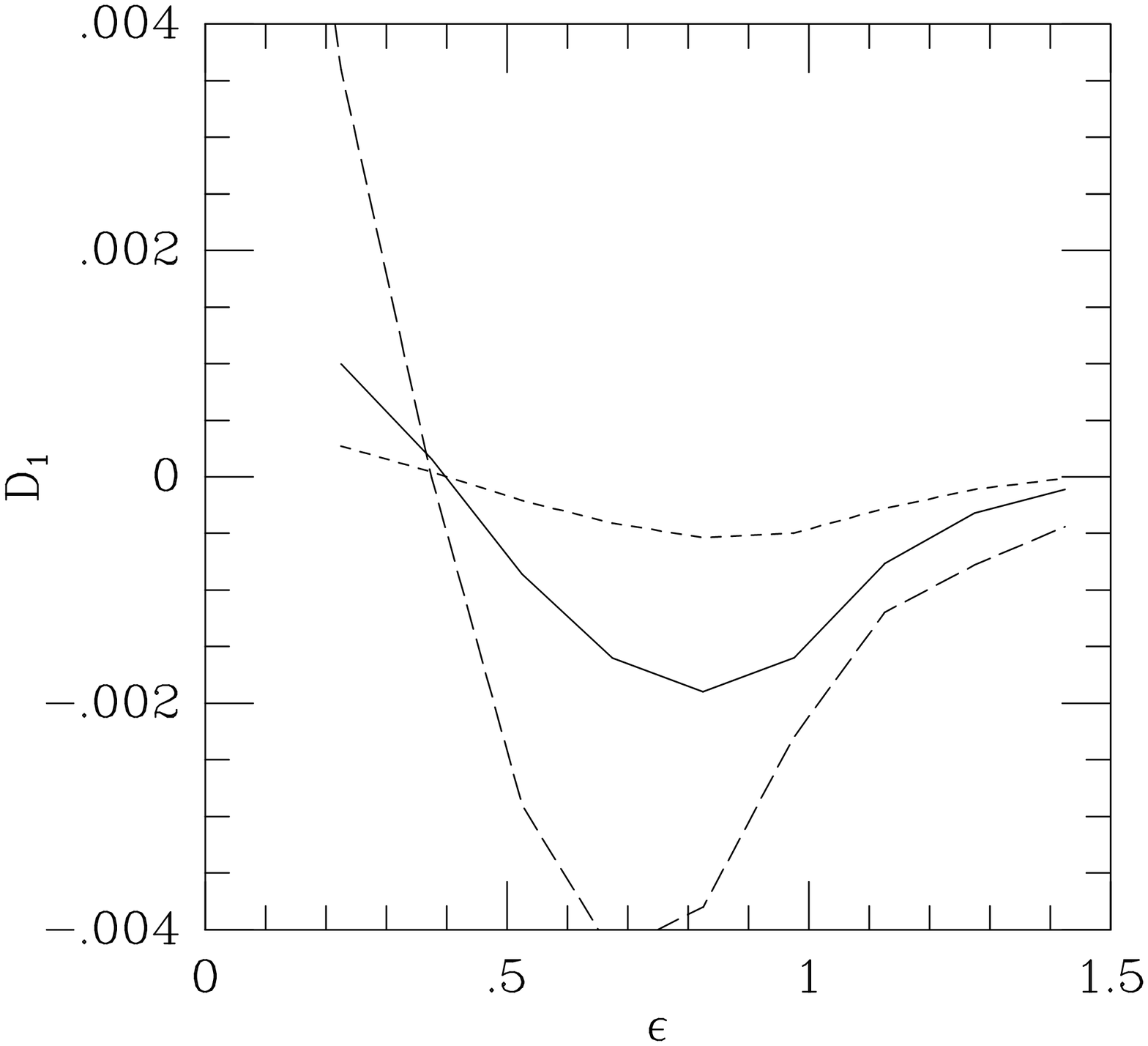} \ \ 
\epsfxsize 7cm
\epsffile{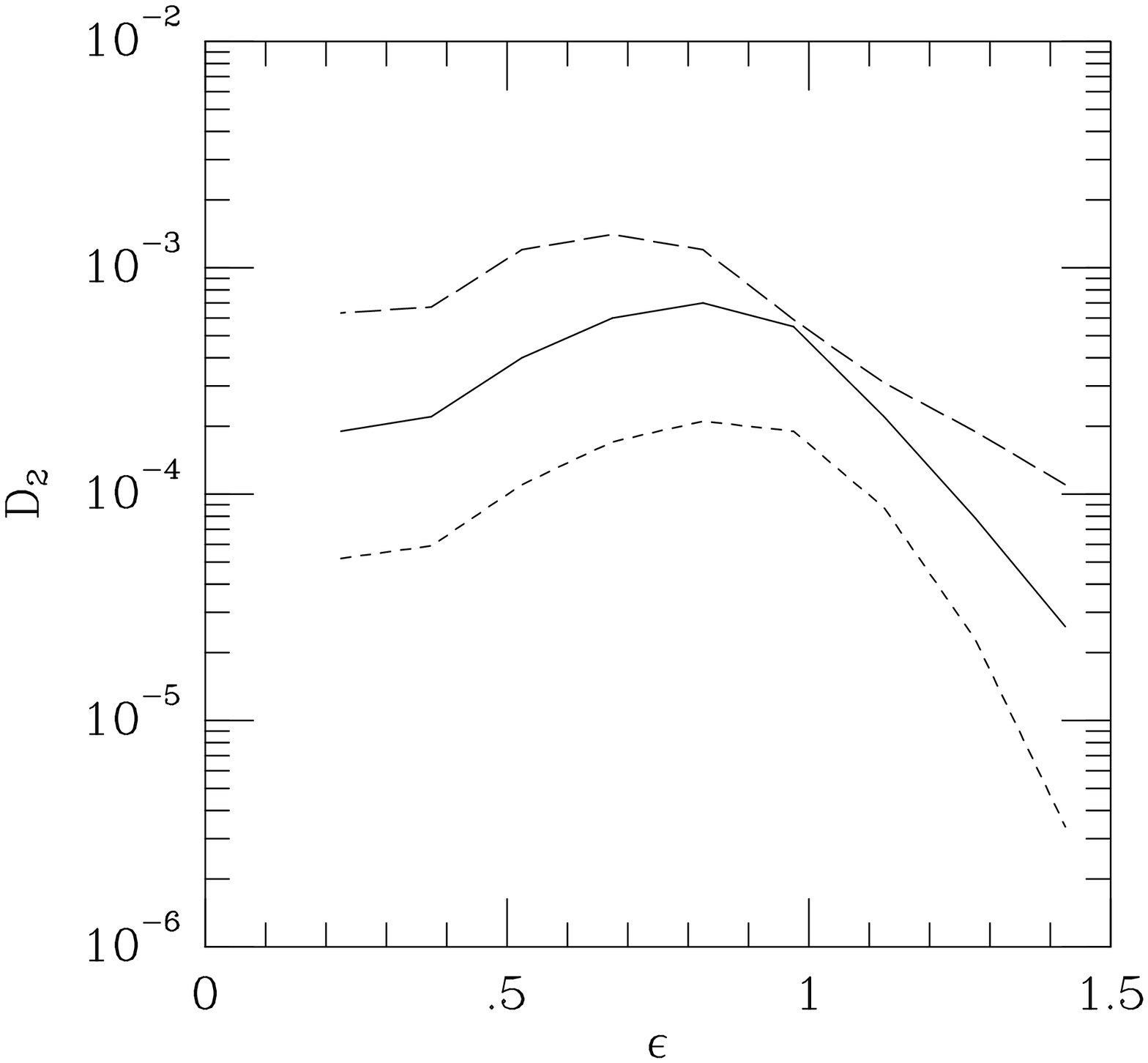}\\
\end{center}
\caption{The diffusion coefficients (a) $D_1$   and (b) $D_2$ plotted against
the energy $e$ for three values of $N$. Long-dashed, solid, and
short-dashed curves are the results for $N=16$, 64 and 256, respectively.}
\label{fig:D}
\end{figure}

In order to see the dependence of these diffusion coefficients on the
energy, we calculated them for intervals of $\Delta \varepsilon = 0.15$. Figure
\ref{fig:D} shows the results, for $N=16, 64$ and 256. The time
interval $\Delta t$ was taken equal to $Nt_c$. We used smaller values
for $\Delta t$ and confirmed that the choice of $\Delta t$ has
negligible effect if $\Delta t$ is larger than $10t_c$ and smaller
than $4Nt_c$.  Time average is taken over the whole simulation period.
We can see that both the first- and second-order terms show very
strong dependence on the energy of the sheets, and of the order of
$1/100N$ for $e \sim 1.5$.  Figure \ref{fig:D} suggests that the
relaxation timescale grows exponentially as energy grows. This
behavior is independent of the value of $N$.

\begin{figure}
\begin{center}
\leavevmode
\epsfxsize 7cm
\epsffile{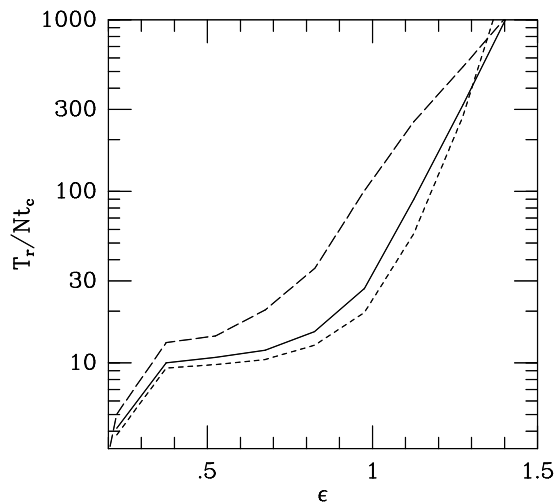}
\end{center}
\caption{The relaxation time in unit of $Nt_c$ plotted against
the energy $e$ for three values of $N$. Curves have the same meanings
as in figure \protect\ref{fig:D}}
\label{fig:tr}
\end{figure}

We can define the relaxation timescale as
\begin{equation}
t_r = {\varepsilon^2/  D_2},
\end{equation}
that is, the timescale in which energy changes significantly. Figure \ref{fig:tr}
shows this relaxation timescale for different values of $N$ and $\varepsilon$. 
The relaxation time shows very strong
dependence on the energy and the relaxation of high-energy sheets is
much slower than that of sheets in lower energies. This is
partly because of the dependence of $t_r$ on $\varepsilon$ itself. However, as w 
can see in figure \ref{fig:D}, the dependence of the diffusion
coefficient is the main reason.

This
result resolves the apparent contradiction between the fact that the
relaxation timescale is of the order of $Nt_c$ \cite{Severne1984}  and 
that the system reaches the true thermal equilibrium only in much
longer timescale.\cite{Tsuchiya1997} \ It is true that the relaxation
timescale is $O(N)$, but the coefficient before $N$ is quite large, in
particular for sheets with high energies. 

An important question is why the relaxation timescale depends so
strongly on the energy. This is provably due to the fact that
high-energy sheets have the orbital period significantly longer than
the crossing time. Typical sheets have the period comparable to the
crossing time, and therefore they are in strong resonance with each
other. However, a high-energy sheet has the period longer than the
crossing time, and thus it is out of resonance with the rest of the
system. Therefore, the coupling between high energy sheets and the
rest of the system is much weaker than the coupling between sheets
with average energy. This explains why the relaxation of high energy
sheets is slow.

\section{Summary and Discussion}

In this paper, we studied the thermal relaxation process in
one-dimensional self-gravitating systems. We confirmed the result
obtained by Tsuchiya {\it et al.}\cite{Tsuchiya1997} that the thermal
relaxation takes place in the timescale much longer than
$Nt_c$. However, we found that this is simply because the thermal
relaxation timescale is much longer than $Nt_c$. Even for typical
sheets, the relaxation timescale is around $10Nt_c$. In order to
obtain good statistics, we need to take average over many relaxation
times. Moreover, the relaxation time for sheets in the high-energy end
of the distribution function is even longer, since the relaxation
timescale grows exponentially as the energy grows. Thus, it is not
surprising that we have to wait for more than $10^4Nt_c$ to obtain good
statistics.

Does this finding have any theoretical/practical relevance? 
Theoretically, there is nothing new in our result. What we found is
simply that numerical simulation should cover the period much longer
than the relaxation timescale to obtain statistical properties of the
system, and that the relaxation timescale of a sheet depends on its
energy. Both are obvious, but some of the previous studies neglected
one or both of the above, and claimed to have found a complex
behavior, which, in our view, is just a random walk.

Our finding of the long relaxation time by itself has rather little
astrophysical significance, since in the large $N$ limit, the
relaxation time is infinite anyway. However, since any numerical
simulation suffers some form of numerical relaxation, it is rather
important to understand how the relaxation effect changes the
system. To illustrate this, we examine the claims by Tsuchiya {\it et
al.}\cite{Tsuchiya1997} in some detail here.

They argued that the evolution of the mass sheet model proceeds in the
following four steps: (1) viliarization, (2) dynamical equilibrium,
(3) quasiequilibrium, and (4) thermal equilibrium. According to them,
the viliarization timescale is order of $t_c$, and the energy of each
sheet is ``conserved'' in the  dynamical equilibrium phase, which
continues up to $t \sim Nt_c$. Then, ``microscopic relaxation'' takes
place in the timescale of $t \sim Nt_c$, where the energy of each
sheet is relaxed, but the whole system needs timescale much longer to 
reach the true equilibrium, because of some complex structure in the
phase space.

Our numerical results are in good agreement with those of Tsuchiya
{\it et al.},\cite{Tsuchiya1997} \ but our interpretation is much
simpler: First system virializes, and then relaxation proceeds in the
timescale of thermal relaxation, which depends on the energy of the
individual sheets. Thus, the central region with short relaxation time 
relaxes to the distribution close to the thermal relaxation in less
than $100Nt_c$, but the distribution in the high-energy tail takes
much longer to settle. In addition, the small number statistics in the 
high-energy region makes it necessary to average over many relaxation
times to obtain good statistics. In other words, there are no
distinction between the ``microscopic'' and ``macroscopic''
relaxation, and the evolution of the system is perfectly understood in 
terms of the standard thermal relaxation.

\section*{Acknowledgments}
I would like to thank Yoko Funato, Toshiyuki Fukushige and Daiichiro
Sugimoto for stimulating discussions, and Shunsuke Hozumi for comments
on the manuscript. This work is supported in part by the Research for
the Future Program of Japan Society for the Promotion of Science
(JSPS-RFTP97P01102).

\end{document}